\documentstyle[twocolumn,amssymb,aps,epsfig,float]{revtex}
\begin{document}
\bibliographystyle{plain}
%\hskip -2\wd0\copy1
\twocolumn[\hsize\textwidth\columnwidth\hsize\csname @twocolumnfalse\endcsname

\title{ 
Residual entropy and spin gap in a one-dimensional
analog of the pyrochlore antiferromagnet
}
\vskip0.5truecm 
\author{M. Mambrini, J. Tr\'ebosc and F. Mila}

\address{Laboratoire de Physique 
Quantique, Universit\'e Paul Sabatier, 118 Route de Narbonne, 31062 
Toulouse Cedex, France. }\vskip0.5truecm
      
\maketitle

\begin{abstract}
\begin{center} 
\parbox{14cm}
{We show that the low-energy
sector of the S=1/2, antiferromagnetic Heisenberg model on 
a one-dimensional lattice of coupled tetrahedra
consists of $2^N$ replica of the spectrum of the 
dimerized Heisenberg chain, where $N$ is the number of tetrahedra. 
This provides a proof of the following properties: i) there is a residual 
ground-state entropy per spin equal to $2^{1/4}$; ii) there is a 
singlet-triplet gap as long as the coupling between the tetrahedra is 
smaller than the internal one. These properties are compared to available
results on the pyrochlore lattice.}
\end{center}
\end{abstract}
\vskip .1truein
 
\vskip2pc
]
%\newpage

It is by now quite clear that the low-energy properties of frustrated magnets
can be very different from those of ordinary magnets with long-range order. In
particular, after the pioneering work of Majumdar and Ghosh on the zigzag 
chain\cite{majumdar}, it has been shown that several systems 
have a singlet-triplet gap in the
magnetic spectrum. This property is actually not specific of frustrated systems
since it is also shared by ladders with an even number of legs\cite{dagotto}. 
There is an
increasing evidence however that frustration can have more specific
consequences, like for instance low-lying singlets in the singlet-triplet gap.
The first example was again the zigzag chain which, contrary to ladders, has a
two-fold degenerate singlet ground-state\cite{majumdar}. 
But the spectrum can be more complex.
For instance, there is a clear numerical evidence that the Heisenberg
antiferromagnet on the Kagome lattice has an exponential number of low
lying singlets below the first triplet excitation\cite{lecheminant,waldtmann}. 
If analytical arguments have
been put forward to explain this property in terms of coupled
triangles\cite{mila}, 
an analytical proof that this is the case is still lacking.
Such properties are actually not limited to low-dimensional systems. In
particular, the Heisenberg model on the pyrochlore lattice is a well known
example of 3D frustrated model without long-range order at the classical
level\cite{classical}, and
the S=1/2, quantum version is believed to have a very short correlation length
and a singlet-triplet gap in the ground-state\cite{canals}.
There is also some indication that there are low-lying singlet states in the
singlet-triplet gap\cite{canals}, 
but again there is no proof that this is indeed the case.

\begin{figure}
\begin{center}
\epsfig{file=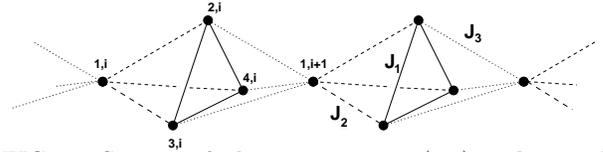,width=7.8cm}
\caption{System of alternating spins $(1,i)$ and triangles
$T_i=\{(2,i),(3,i),(4,i)\}$. The notations for the couplings are: $J_1$ for
bonds inside triangles and  $J_2$ (respectively $J_3$) between a triangle $T_i$
and its left (right) spin neighbor $S_{1,i}$ ($S_{1,i+1}$).}
\label{fig:chain}
\end{center}
\end{figure}

In this paper, we show that these properties actually occur in a
one-dimensional analog of the pyrochlore lattice, namely a one-dimensional
array of coupled tetrahedra. The discussion can actually be carried out for the
slightly more general situation of a one dimensional system of alternating
spins and triangles (see Fig. \ref{fig:chain}) defined by the Hamiltonian:

\begin{eqnarray}
{\cal H} & = & J_1 \sum_i (\vec S_{2,i}.\vec S_{3,i}+\vec S_{3,i}.\vec S_{4,i}+
\vec S_{4,i}.\vec S_{2,i}) \nonumber \\
& + & J_2 \sum_i \vec S_{1,i}.(\vec S_{2,i}+\vec S_{3,i}+\vec S_{4,i}) 
\nonumber \\
& + & J_3 \sum_i (\vec S_{2,i}+\vec S_{3,i}+\vec S_{4,i}).\vec S_{1,i+1}
\label{hamiltonian}
\end{eqnarray}
where $\vec S_{n,i}$ are spin 1/2 operators. Similar models involving more than
one spin at a given site have been studied\cite{kolezhuk,niggemann,takano,tonegawa}, but to our knowledge none of them
dealt with triangles, and the properties that are described below are
very specific to that case.

%%%
%% FIG 1
%%%

The most important step toward a solution of this Hamiltonian is to realize
that the spins of the triangle of any unit cell $i$ enter the Hamiltonian only 
through their sum $\vec T_i=\vec S_{2,i}+\vec S_{3,i}+\vec S_{4,i}$ since the
first term can be rewritten $(J_1/2)\sum_i(\vec T_i^2-9/4)$. This observation
has two consequences. First of all, it means that the
total spin of the triangle of any unit cell, which can take the degenerate value
3/2 or the twofold degenerate value 1/2, is a good quantum number. Secondly, 
it shows that the additional quantum number needed to specify the states 
in the subspace of total spin 1/2 - for instance the chirality - does 
not enter the Hamiltonian. 

So the eigenvalues of the problem of Eq. (\ref{hamiltonian}) are the same as
those of the following family of effective 
Hamiltonians ${\cal H}(\{T_i\})$:
\begin{eqnarray}
{\cal H}(\{T_i\}) & = & \sum_i(J_2\, \vec S_{1,i}. \vec T_i + J_3\, \vec T_i . \vec
S_{1,i+1}) \nonumber \\
& + & (J_1/2)
\sum_i(\vec T_i^2-9/4) 
\label{effective}
\end{eqnarray}
where $\vec T_i^2=T_i(T_i+1)$ and $T_i=1/2$ or $3/2$, 
the degeneracy of the eigenvalues of ${\cal H}(\{T_i\})$
for the original problem being equal to $2^{n_{1/2}}$, where $n_{1/2}$ is the
number of triangles with total spin 1/2.

The twofold degeneracy associated with each triangle in a doublet state can
actually be understood in a more direct way. Let us consider the eigenstates of
the problem obtained by considering only one spin in the triangle of 
unit cell $i$, say $\vec S_{2,i}$. Then the wave functions obtained  
as the product of the
eigenstates of that new problem with the singlet constructed out of 
$\vec S_{3,i}$ and $\vec S_{4,i}$ are eigenstates of the original problem since
all the extra couplings connect a single spin to both ends of that singlet.
Besides, $\vec S_{3,i}$ and $\vec S_{4,i}$ forming a singlet, these eigenstates
correspond to $T_i=1/2$. Finally, the wave functions obtained by putting the
singlet on the three possible bonds of one triangle are not linearly 
independent but generate a space of dimension 2, as for a single triangle.

So the problem has now been split into different sectors corresponding to the
values of the $T_i$'s. For clarity, we will continue the discussion in two
specific situations: i) Alternating spins 1/2 and
triangles, where the analysis is particularly straightforward; 
ii) Coupled tetrahedra,
which is physically more relevant as an analog of the pyrochlore
antiferromagnet.

\section{Alternating spins 1/2 and triangles} 

\begin{figure}
\begin{center}
\epsfig{file=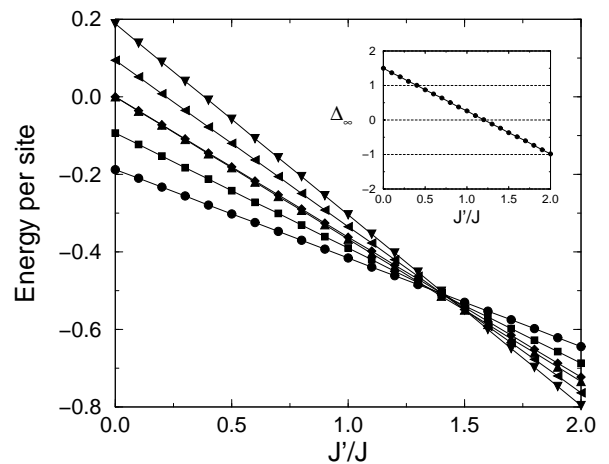,width=7.8cm}
\caption{{\it Main figure}. Ground state energy per site of ${\cal H}(\{T_{i \neq
i_0}=1/2,T_{i_0}=3/2\})$ for a $L=16$ cluster (i.e. 4 tetrahedra) for 
different values of the number of triangles $T_i=3/2$: ($\bullet$) $n_{3/2}=0$, 
($\blacksquare$) $n_{3/2}=1$, ($\blacktriangle$) $n_{3/2}=2$ neighboring triangles, 
($\blacklozenge$) $n_{3/2}=2$ next-neighboring triangles, ($\blacktriangleleft$)
$n_{3/2}=3$, ($\blacktriangledown$) $n_{3/2}=4$.
{\it Inset}. Thermodynamic limit of the energy difference between the GS of 
${\cal H}(\{T_{i \neq i_0}=1/2,T_{i_0}=3/2\})$ and
${\cal H}(\{T_{i}=1/2\})$. Up to $J'_{c} \simeq 1.2 J$ the low energy physics
of the model is given by the spin 1/2 Heisenberg model.}
\label{fig:scal1}
\end{center}
\end{figure}

This case corresponds to $J_2=J_3$. For clarity, we rename the parameters
$J_1=J$ and $J_2=J_3=J'$. So we are dealing in this section with triangles of
strength $J$ coupled to spins 1/2 with 3 exchange integrals of strength $J'$.
In the limit where $J'/J$ goes to zero,
the spectrum of each sector {$T_i$} is completely degenerate, and the energy
is given by the sum of the energies of the triangles
$E(\{T_i\})=-(3/2)(N-\sum_i T_i)$,
where N is the number of unit cells. In
that limit, the lowest energy states are those obtained when all the triangles
are in a doublet. Perturbative arguments show that this will remain true up 
to a certain value of the ratio
$J'/J$. To study that problem quantitatively, we have 
performed exact
diagonalizations of finite clusters for the Hamiltonians of Eq. (\ref{effective}).  
The results are given in Fig. \ref{fig:scal1}.

{\it Finite size analysis.} It is well known 
that the ground state energy per site of the spin 1/2 Heisenberg
model scales like \hbox{$e_{L}=e_{\infty}-A/L^{2}$} where
\hbox{$e_{\infty}=1/4-\ln 2$}. So the ground state energy per site in units
of $J$ for ${\cal H}(\{T_i=1/2\})$, which is nothing but the standard Heisenberg
model with coupling $J'$ up to a constant of order $J$, scales like
\hbox{$\varepsilon^{(0)}_L=-3/16+(J'/2J)e_{\infty}-A/L^{2}$} where $L=4N$ is
the total number of sites of the system. Let us focus now on
\hbox{$\varepsilon^{(1)}_L$}, the GS per site of 
\hbox{${\cal H}(\{T_{i \neq i_0}=1/2,T_{i_0}=3/2\})$}, and let us denote by 
$\delta_{\infty}$ the thermodynamic limit of the energy difference between the 
GS energy of the spin 1/2 Heisenberg model and the GS energy of the same model 
with one spin 1/2 replaced by a spin 3/2. 
Since there are no $1/L$
corrections for 
$\varepsilon^{(0)}_L$, one should expect the following scaling
for \hbox{$\varepsilon^{(1)}_L$},

\begin{equation}
\varepsilon^{(1)}_L=\varepsilon^{(0)}_{\infty}+
\Delta_{\infty}/L+{\cal O}(1/L^2)
\label{trianglescale}
\end{equation}
where $\Delta_{\infty}=(3-(J'/J)\delta_{\infty})/2$ is the thermodynamic limit of the energy difference between the GS of 
${\cal H}(\{T_{i \neq i_0}=1/2,T_{i_0}=3/2\})$ and
${\cal H}(\{T_{i}=1/2\})$.
Numerical simulations up to $L=28$ with one spin $3/2$ show that the 
scaling (\ref{trianglescale}) is very well verified and allowed us to extract
the gap (see Fig. \ref{fig:scal1}) which, as expected, is linear in $J'$,
and  $\delta_{\infty} \simeq 5/2$.

%%%
%% FIG 2
%%%

{\it Discussion.} It is then clear that the
ground state remains in the sector \{$T_i=1/2$\} up to $J'_c \simeq 1.2 J$, and in particular for the isotropic point $J'=J$. When this is the case,
the low energy physics is given by the Hamiltonian of Eq. (\ref{effective}) with
all $T_i$'s equal 1/2, which, as we saw, is nothing but the one-dimensional, 
spin 1/2 
Heisenberg model with coupling $J'$. The only difference is that we now deal
with $2^{N}$ replica of the spectrum. All properties of interest can be deduced
from this mapping. In particular, there is no singlet-triplet gap in the
spectrum, the elementary excitations are deconfined spinons, and there is a 
residual entropy per spin equal to $2^{1/4}$.

\section{Coupled tetrahedra}

This case corresponds to $J_2=J_1$. For clarity we use the notations
$J_2=J_1=J$ and $J_3=J'$ in the following. 
So we are now dealing with tetrahedra of strength $J$
coupled by bonds of strength $J'$. Let us start from the limit $J'=0$. In that
case, the groundstate is obtained by putting each tetrahedron in its
groundstate. Since the ground state of a tetrahedron is twofold degenerate
with energy
$-(3/2)J$, the groundstate is $2^{N}$-fold degenerate with energy $-(3/2)NJ$.
The low lying excited states correspond to putting one tetrahedron in its first
excited state with an energy cost of $J$. With respect to the general
classification of the states proposed at the beginning of the paper, the
situation is slightly more complicated than in the previous case. First of all,
the groundstate manifold in the limit $J'=0$, of dimension $2^{N}$, contains 
only some of the eigenstates of ${\cal H}(\{T_i=1/2\})$, the Hilbert space of that 
Hamiltonian being of dimension $2^{2N}$. 
Besides, the manifold of the first excitation of energy $J$
contains states corresponding to
different effective Hamiltonians. This comes from the fact that the triplet
excitation of a tetrahedron is threefold degenerate, and that only two of them
can be constructed with a given triangle being a doublet, the other one
corresponding to the groundstate of a spin 1/2 coupled to the spin 3/2 state of
a triangle.
So it is no longer clear a priori that the low
energy sector can be described by only one of the effective Hamiltonians of Eq.
(\ref{effective}). 

%%%
%% FIG 2
%%%
\begin{figure}
\begin{center}
\epsfig{file=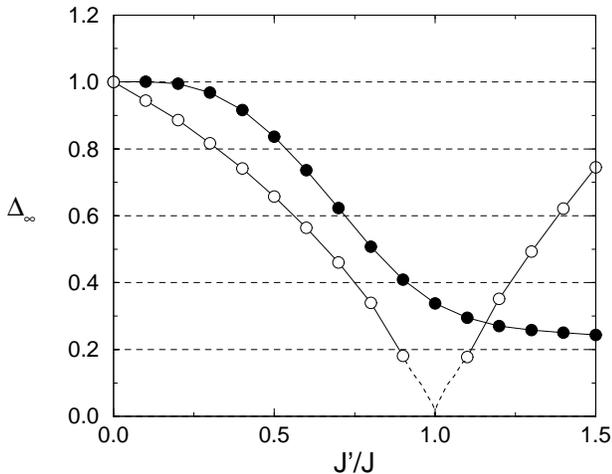,width=7.8cm}
\caption{Extrapolated values of the GS energy difference between
${\cal H}(\{T_{i \neq i_0}=1/2,T_{i_0}=3/2\})$ and ${\cal H}(\{T_{i}=1/2\})$ ($\bullet$)
and singlet-triplet gap of ${\cal H}(\{T_{i}=1/2\})$ ($\circ$). Up to large
values of $J'$ including $J'=J$, the effective hamiltonian of the system is
${\cal H}(\{T_{i}=1/2\})$ and the low energy properties are those of the spin 1/2 dimerized Heisenberg model.}
\label{fig:scal2}
\end{center}
\end{figure}

To address this point, we have again resorted to exact
diagonalizations of finite clusters, and we have determined which effective
Hamiltonian gives the groundstate and the first triplet excited state as a
function of $J'$. The results are given in Fig. \ref{fig:scal2}. As in the previous case, the 
groundstate is given by ${\cal H}(\{T_i=1/2\})$ regardless of the size of the system up
to very large values of $J'$, and in
particular beyond the isotropic limit $J'=J$. 
For the first excited state, the situation is slightly more involved since 
${\cal H}(\{T_i=1/2\})$ is now equivalent to the dimerized Heisenberg model and 
has a gap if $J'\ne J$\cite{cross}. For small
enough $J'$, the first excited state is also given by ${\cal H}(\{T_i=1/2\})$ for all 
sizes. However, this is no longer true if $J'$ is beyond a certain value
$J'_c(N)$ which
increases with the size of the system. Since we are interested in the properties
of the system in the thermodynamic limit, we have performed a finite size
analysis of the first excited state of ${\cal H}(\{T_{i}=1/2\})$ and of the
GS of ${\cal H}(\{T_{i \neq i_0}=1/2,T_{i_0}=3/2\})$ (see results in Fig. \ref{fig:scal2}).

{\it Finite size analysis.} Contrary to the situation discussed in the 
previous section,
there are exponential corrections to the finite size
energies if $J'\ne J$ because there is then gap in the spectrum of 
${\cal H}(\{T_i=1/2\})$.
Namely, with the same notations as in section I,
\begin{equation}
\varepsilon^{(1)}_L=\varepsilon^{(0)}_{\infty}+
\Delta_{\infty}/L-(A/L^2) e^{-L/\xi}
\label{tetrascale}
\end{equation}
Such a scaling for clusters up to $L=28$ with one spin 3/2 in the system
allows the determination of \hbox{$\Delta_{\infty}(\{T_{i \neq i_0}=1/2,T_{i_0}=3/2\})$} which remains positive up
to very large values of $J'$ (at least~$2J$). 
Let us turn to the case of the first excited state of ${\cal H}(\{T_{i}=1/2\})$.
Again, a scaling of the type 
\hbox{$\Delta_{L}=\Delta_{\infty}+(A/L)e^{-L/\xi}$} 
is expected with $\xi$ diverging and
 \hbox{$\Delta_{\infty}(\{T_{i}=1/2\})=0$}
at the isotropic point
$J'=J$. Since the systems contains only spin 1/2 is that sector we performed
diagonalizations up to $L=48$ to calculate $\Delta_{\infty}$. 
The results (see Fig. \ref{fig:scal2}) clearly agree with what was expected 
even if the precise behavior of $\Delta_{\infty}\sim \delta^{2/3}$\cite{cross} 
(up to logarithmic corrections) as the dimerization
\hbox{$\delta=(1-J'/J)/(1+J'/J)$} goes to
zero is difficult to extract from numerical simulations. 

{\it Discussion.} The results are quite clear: For large systems, $J'_c$ is
larger than $J$, namely $J'_c \simeq 1.15 J$. Then to understand the low 
energy properties of the model in the parameter range
$J'\le J$ and in the thermodynamic limit, one can use ${\cal H}(\{T_i=1/2\})$ as an 
effective Hamiltonian. This means that the low energy physics can be described 
by
the dimerized, spin 1/2 Heisenberg chain with
alternating exchange integrals $J$ and $J'$. This 
model has been extensively
studied in the context of spin-Peierls systems\cite{cross}. The main property,
already used to perform the finite-scaling,
is that there is a
singlet-triplet gap in the system as long as $J'<J$, and that this energy gap 
closes at the point $J'=J$. Besides, the elementary excitations are boundstates 
of spinons due to the confinement potential introduced by the dimerization \cite{affleck}.
Regarding the properties of the original problem,
one should not forget that all these states are degenerate, and in particular
that there is, as in the previous case, a residual entropy per spin equal to 
$2^{1/4}$.

\section{Discussion}

The main motivation in undertaking the present study was to shed some light on
the properties of the pyrochlore antiferromagnet. This system is a three
dimensional structure that can be thought of as an array of coupled tetrahedra.
The current situation as far as a theoretical understanding of that model is
concerned is the following: There is a clear evidence that the spin-spin
correlation functions are extremely short ranged, and there is some numerical
evidence that there is a singlet-triplet gap in the spectrum \cite{canals}. Both properties
are indeed satisfied by the present 1D model of coupled tetrahedra in the
parameter range $J'<J$. In particular, since the ground states can be written as
products of local singlets, the spin-spin correlation functions indeed decay
very fast. The fact that the gap closes when $J'=J$ is not really inconsistent
with the case of the pyrochlore since there is an important difference in the
way tetrahedra are coupled in both cases:
Pairs of tetrahedra are never coupled through more than one spin on each of them 
in the pyrochlore structure whereas one of the tetrahedra is coupled through 3
spins in the present model. So $J'=J$ in the present case
somewhat corresponds to a stronger coupling than for the pyrochlore. 

More importantly, the presence of a residual entropy per spin in the model of
the present paper suggests that, if there is a
singlet-triplet gap in the spectrum, there should indeed 
be low-lying singlets within this gap. This should be particularly easy to
detect if one goes away from the standard pyrochlore and considers a dimerized
version of the same model with weakly coupled tetrahedra since the gap becomes
larger in that limit while at the same time the splitting between the singlet 
states decreases. Work is in progress along these lines.

To summarize, we have proposed and solved a one dimensional analog of the
pyrochlore antiferromagnet
and proved that it exhibits at least some 
of the exotic physics one can hope to find in very frustrated magnets,
namely: A singlet-triplet gap, and a lot of low-lying singlets. 
Given the relative simplicity of the model, it is the authors hope that some
compound can be synthesized with this kind of structure. In any case, the very
simple picture of quite unusual properties provided by this model is expected to
serve as a useful guide in the search of new experimental realizations of very
frustrated magnets.

%\begin{figure}[hp]
%\centerline{\psfig{figure=fig1.eps,width=6.0cm,angle=0}}
%\vspace{0.5cm}
%\caption{Sketch of the dimerized Kagome lattice. Solid lines: $J$,
%dashed lines: $J'$.}
%\label{fig1}
%\end{figure}

The numerical simulations were performed on the Cray supercomputers 
of the IDRIS (Orsay, France).


\begin{references}

\vspace{-1.cm}
\bibitem{majumdar} C. K. Majumdar and D. Ghosh, J. Math. Phys. {\bf 10}, 1388
(1969).

\bibitem{dagotto} For a review, see E. Dagotto and T. M. Rice, Science {\bf
271}, 618 (1996).

\bibitem{lecheminant} P. Lecheminant, B. Bernu, C. Lhuillier, L. Pierre and P.
Sindzingre, Phys. Rev. B {\bf 56}, 2521 (1997).

\bibitem{waldtmann} C. Waldtmann, H.-U. Everts, B. Bernu, 
C. Lhuillier, P. Sindzingre, P. Lecheminant, L. Pierre, Eur. Phys. J. B {\bf 2},
501 (1998).

\bibitem{mila} F. Mila, Phys. Rev. Lett. {\bf 81}, 2356 (1998).

\bibitem{classical} A. V. Chubukov, Phys. Rev. Lett. {\bf 69}, 832 (1992); A. B.
Harris, C. Kallin and A. J. Berlinsky, Phys. Rev. B {\bf 45}, 2899 (1992); P.
Chandra, P. Coleman and I. Ritchey, J. Phys. I (France) {\bf 3}, 591 (1993).

\bibitem{canals} B. Canals and C. Lacroix, Phys. Rev. Lett. {\bf 80}, 2933  
(1998).

\bibitem{kolezhuk}  A. K. Kolezhuk and H.-J. Mikeska, Phys. Rev. B {\bf 56},
R11380 (1997).

\bibitem{niggemann} H. Niggemann, G. Uimin and J. Zittartz,
J. Phys.: Condens. Matter
{\bf 9} (1997) 9031-9042.

\bibitem{takano} K. Takano, K. Kubo, H. Sakamoto, J. Phys.: Condens. Matter
{\bf 8} (1996) 6405-6411.

\bibitem{tonegawa}  T. Tonegawa, T.
Hikihara, T. Nishino, M. Kaburagi, S. Miyashita, H.-J. Mikeska, 
J. Magn. Magn. Matter. (Proc. Int. Conf. Magnetism, Cairns, 1997)

\bibitem{cross} M. C. Cross and D. S. Fisher, Phys. Rev. B {\bf 19}, 402 (1979).

\bibitem{affleck} I. Affleck, \textit{Dynamical properties of
unconventional magnetic systems} (A. T. Skjeltorp and
D. Sherrington, Kluwer Academic, Dordrecht, Boston 1998).


\end{references}
\end{document}